\documentclass[fleqn,usenatbib]{mnras}

\usepackage{newtxtext,newtxmath}
\usepackage[T1]{fontenc}
\usepackage{graphicx}	
\usepackage{amsmath}	
\usepackage{booktabs}
\usepackage{hyperref}
\usepackage{appendix}
\title[PINN Analysis of Cloud Stability in TGQ]{
A Physics Informed Neural Network Approach to
Decipher Environmental Confinement of Molecular Clouds in the Third Galactic Quadrant}

\author[Mondal et al.]{
Ashok Mondal,$^{1}$\thanks{E-mail: ashok.contai@gmail.com}
\\
$^{1}$Department of Applied Science and Humanities, Haldia Institute of Technology, West Bengal, India
}

\date{Accepted XXX. Received YYY; in original form ZZZ}
\pubyear{2024}

\begin{document}
\label{firstpage}
\pagerange{\pageref{firstpage}--\pageref{lastpage}}
\maketitle

\begin{abstract}
Recent high-sensitivity and high-resolution surveys of the Third Galactic Quadrant [$219.75 \leq l \leq 229.75$ and $-5.25 \leq b \leq 5.25$], such as the Milky Way Imaging Scroll Painting ( MWISP) survey, have revealed a significant population of molecular clouds (MCs) that appear virially unbound ($\alpha_{vir} > 2$) according to traditional isolated MC models. However, these apparently unstable cloud structures are actually found  long-lived and actively undertaking star formation. This inconsistency presents a Virial Paradox.  In the present work, we employ a Physics-Informed Neural Network (PINN) to address this discrepancy by  re-evaluating the stability of 2,091 MCs . Unlike traditional machine learning, our PINN model embeds the complete Eulerian Virial Theorem including external surface pressure support ($P_{ext}$) directly into the network's loss function. This model autonomously discovers a global Galactic pressure field that scales inversely with vertical height ($|z|$) and Galactocentric radius ($R_{GC}$), consistent with the properties of the Galactic Flare. By incorporating this AI-derived pressure support, we demonstrate a dramatic shift in the stability census: 99.9\% of previously thought unbound clouds are found to be in a state of pressure confined equilibrium ($\alpha_{AI} \leq 2$). Furthermore, we find a robust log-log correlation ($R = 0.64$) between our corrected stability parameter and the star formation rate (SFR). This correlation  provides independent physical validation that environmental confinement is a primary regulator of star forming efficiency in the outer Galaxy.
\end{abstract}

\begin{keywords}
ISM: clouds -- stars: formation -- Galaxy: structure -- methods: numerical -- HII regions
\end{keywords}


\section{Introduction}

The stability of molecular clouds is a fundamental prerequisite for star formation, which requires a delicate balance between outward internal support and inward gravitational forces [\citealt{McKee_2007}; \citealt{Chevance_2020}]. Gravitational boundedness has traditionally been represented by the virial parameter, $\alpha_{vir} = 2K/|\Omega|$, where $K$ and $\Omega$ represent the internal kinetic energy and  the gravitational potential energy respectively.
When $\alpha_{vir} < 2$, MCs are considered to be gravitationally bound ( \citealt{Bertoldi_1992}, \citealt{Kauffmann_2013} ).
However, the high resolution datasets of \citet{Dong_2023} indicate that a vast majority of bound clouds in the Third Galactic Quadrant (TGQ) appear significantly unbound as their virial parameter values obtained from the observed velocities and masses is higher ($\alpha_{vir} \gg 2$).  This  physical discrepancy is usually known as the Virial Paradox of MCs. These clouds having higher $\alpha_{vir}$ should be rapidly dispersing into the interstellar medium (ISM), However, observational evidences (\citealt{Yuehui_2022},\citealt{Wang_2023}, \citealt{Dong_2023}) confirm that these structures are not only long-lived but are also the primary sites of active star formation. \citet{Ashok_2024} established that star formation rates (SFR) are intimately tied to cloud properties; however, if these clouds were truly unbound, they would disperse before star formation could proceed. This suggests a missing component in traditional virial analysis: the role of the surrounding galactic environment.

\noindent
Molecular clouds are not isolated systems existing in a vacuum, rather they are embedded within a high pressure Interstellar medium that exerts a confining force on their boundaries. To resolve the above mentioned virial paradox, we expanded the classical virial equation to include an external surface pressure term, $P_{ext}$. With this $P_{ext}$ we have considered the energy balance as:
\begin{equation}
    2K + \Omega - 4\pi R^3 P_{ext} = 0.
    \label{Equation: Pressure balance}
\end{equation}
\noindent
Where $K$ is the total outward internal kinetic energy and $\Omega$ is the inward gravitational potential energy.
Assuming MCs are spherical with uniform density distribution, we have calculated the values of $K$ and $\Omega$ as:
 
\begin{equation}
    K = \frac{3}{2} M \sigma_{v}^2 \quad \text{and} \quad \Omega = -0.6 \frac{GM^2}{R}.
\end{equation}
Where $M$ is the cloud mass, $R$ is the effective cloud radius, and the term $\sigma_{v}$ represents the one dimensional velocity dispersion, derived here from the standard deviation of the local standard of rest velocity ($V_{LSR}$) line profile as reported in the \citet{Dong_2023} catalog. 
Although, the Eq. \ref{Equation: Pressure balance} can theoretically explain the instability,the $P_{ext}$ can not be observed directly and remains challenging to estimate ( \citealt{Field_2011}; \citealt{Kauffmann_2013}, \citealt{Chevance_2020}). This is because it depends on several complex Galactic environmental factors, such as the vertical distance from the Galactic plane ($|z|$) and the Galacto centric radius $R_{GC}$.

\noindent
In this study,  we apply a Physics Informed Neural Network (PINN) to predict $P_{ext}$. Traditional neural networks learn patterns solely from data, whereas, a PINN incorporates the underlying physical laws directly into the learning process. In our model, the modfied Virial Equation (Eq. \ref{Equation: Pressure balance}) is embedded in the model's loss function. This ensures that the model's prediction of $P_{ext}$ remains consistent with the fundamental principles governing spherical MC dynamics. Using this physics informed framework our  PINN model is trained to infer the external pressure  that satisfies the virial equilibrium condition for each cloud. We apply this approach to a sample of 2,091 MCs that exhibit virial parameters significantly larger than the classical gravitationally bound threshold as mentioned by \citet{Dong_2023}.
Inferred $P_{ext}$ helps us to test whether these apparently unbound MCs can remain stable. We also have compared the pressure induced dynamical state of a MC with it's observed star forming activity. We have examined whether the predicted external pressure can explain why these clouds survive for a long time and continue to form stars despite having large virial parameter values.  This study provides a new inverse modelling approach to understand cloud stability and find a answer of the Virial Paradox.

\noindent
In Section \ref{Data Source and Preprocessing}, the mathematical model is described. Following this, Section \ref{Methodology} outlines and deliberates upon the methodology. Section \ref{Results and Discussion} encompasses the presentation and discourse of results, while the concluding remarks can be found in Section \ref{Conclusion}.

\section{Data Source and Preprocessing}
\label{Data Source and Preprocessing}
The molecular cloud (MC) catalog utilized in this study is derived from the large scale survey conducted by \cite{Dong_2023}, focusing on the Third Galactic Quadrant (TGQ). This dataset provides a multi tracer inventory of the interstellar medium, categorizing structures based on isotopic tracers of varying densities. The catalog specifically includes:
\begin{itemize}
    \item \textbf{1,502 $^{12}$CO structures:} Representing the low-density, diffuse envelopes of the molecular complexes.
    \item \textbf{570 $^{13}$CO structures:} Identifying intermediate density regions.
    \item \textbf{53 C$^{18}$O structures:} Corresponding to high density regions and dense cores, which are the primary sites of gravitational collapse.
\end{itemize}

To ensure the numerical stability and physical validity of the PINN training, a rigorous preprocessing pipeline is implemented. We have performed list-wise deletion of entries containing null values or empty strings. Furthermore, we excluded any structures with non-physical parameters, specifically filtering for rows where Mass $> 0$ and Radius $> 0$. Following these quality control measures, the final analysis sample consists of 2,091 clouds, providing a robust statistical cross-section of the TGQ molecular population.

To account for the clouds' positioning within the Galactic potential and the resulting environmental pressure, observational coordinates (Galactic Longitude $l$, Latitude $b$, and Distance $D$) were transformed into physical Galactic coordinates. We calculated the vertical height ($z_{pc}$) relative to the midplane and the galactocentric radius ($R_{GC}$), assuming a solar distance of $R_{\odot} = 8.15$~kpc:
\begin{equation}
    z_{pc} = D \cdot 1000 \cdot \sin(\text{radians}(b))
\end{equation}
\begin{equation}
    R_{GC} = \sqrt{R_{\odot}^2 + D^2 - 2R_{\odot}D \cdot \cos(\text{radians}(l))}
\end{equation}
\noindent
These derived spatial coordinates, along with Mass ($M_{\odot}$), Radius (pc), and Velocity Dispersion ($\sigma_{v}$), were used as the five primary input features for the neural network. For the physical constraints in the loss function, we utilized the tracer-specific Flag (where Flag = 18 denotes C$^{18}$O) to impose stricter gravitational boundedness requirements on the dense core sub-population.

\section{Methodology}
\label{Methodology}
The standard virial analysis of molecular clouds typically defines the stability of a structure via the virial parameter $\alpha_{vir} = M_{vir}/M_{obs}$. This approach assumes that clouds are isolated, uniform, and non-magnetized spheres. However, as noted in the recent catalog of the Third Galactic Quadrant (TGQ) by \citet{Dong_2023}, a substantial majority of star-forming clouds appear significantly unbound ($\alpha_{vir} \gg 2$). 
This discrepancy suggests that internal gravity alone is insufficient to explain the observed lifetimes of these structures. To resolve this virial paradox, we propose that environmental confinement via external pressure ($P_{ext}$) is a non-negligible factor. To account for this environmental support, we define the AI-corrected virial parameter, $\alpha_{AI}$, as:
\begin{equation}
    \alpha_{AI} = \frac{2K}{|\Omega| + 4\pi R^3 P_{ext}}
\end{equation}
where $P_{ext}$ is the external pressure discovered by the PINN. Values of $\alpha_{AI} \leq 2$ indicate a structure that has successfully reached equilibrium through a combination of self-gravity and environmental confinement. Since $P_{ext}$ cannot be measured directly through CO observations, we employ a PINN to derive the pressure fields required to maintain Galactic cloud stability.

\subsection{The PINN Architecture}
The PINN architecture is implemented as a deep Multi-Layer Perceptron (MLP) consisting of an input layer, three hidden layers, and an output layer. The specific configuration is detailed below:

\begin{itemize}
    \item \textbf{Input Layer:} Comprises 5 neurons corresponding to the physical and environmental parameters: $\{M, R, \sigma_v, z, R_{GC}\}$. To ensure numerical stability and prevent gradient saturation, all inputs are normalized to zero mean and unit variance.
    
    \item \textbf{Hidden Layers:} 
    We utilize three hidden layers, each consisting of 64 fully-connected neurons. 
Three hidden layers provide the necessary depth to model the composite non linearities of the Galactic pressure field. According to the Universal Approximation Theorem, this depth allows for hierarchical feature extraction, moving from raw physical parameters to complex thermodynamic states.
   A width of 64 neurons per layer was selected to provide sufficient expressive power to capture large scale Galactic trends (e.g., the Galactic Flare) while maintaining a parameter-to-sample ratio that prevents overfitting. With approximately 8,641 trainable parameters against 2,091 data points, the model maintains a high degree of generalization.
The 64 neuron width leverages hardware level optimizations for power of two matrix operations, ensuring rapid convergence during the 4,000 training epochs without sacrificing numerical precision in the physics-informed loss function.

    The \textit{Tanh} activation function was specifically selected for these layers to ensure $C^{\infty}$ smoothness throughout the network. Unlike the non-differentiable nature of ReLU at zero, the smooth derivatives of \textit{Tanh} are essential for the backpropagation of gradients through the physics-informed loss function.
    
    \item \textbf{Output Layer:} A single neuron represents the predicted external pressure $P_{ext}$. We apply a \textit{Softplus} activation function to the output layer:
    \begin{equation}
    \text{Softplus}(y) = \ln(1 + e^y)
\end{equation}
The use of Softplus is critical as it enforces the physical boundary condition that external pressure must be positive ($P_{ext} > 0$), a constraint that standard linear or ReLU layers cannot guarantee.
    
\end{itemize}

The network is trained using the Adam optimizer with a learning rate of $\eta = 10^{-4}$. The total number of trainable parameters is approximately 8,641, which provides a robust fit for the 2,091 data points available in the TGQ catalog.

\begin{figure*}
    \centering
    \includegraphics[width=0.85\linewidth]{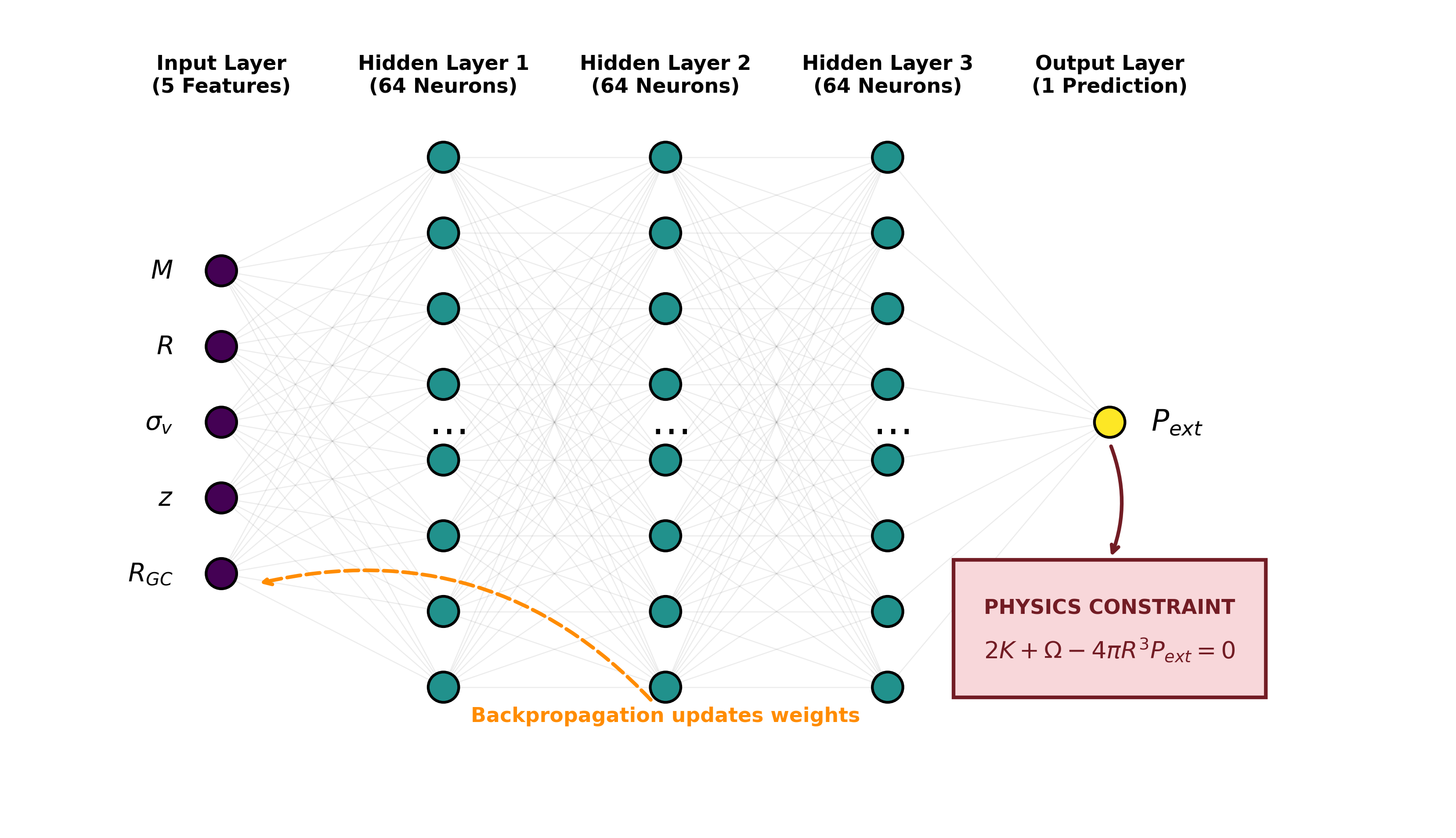}
    \caption{Architecture of the Physics-Informed Neural Network (PINN) used to derive environmental pressure ($P_{ext}$) for the sample of 2,091 molecular clouds. The model processes five physical inputs through three hidden layers (64 neurons each) to predict a single scalar output. The unique nature of the PINN is highlighted by the \textit{Physics Constraint} block; the predicted $P_{ext}$ is fed into the Virial Equation to calculate a physical residual, which is then backpropagated to update the network weights, ensuring the discovered pressure is consistent with gravitational equilibrium.}
    \label{fig:pinn_architecture}
\end{figure*}

The structural framework of our Physics-Informed Neural Network is illustrated in Figure~\ref{fig:pinn_architecture}. We employ a deep Multi-Layer Perceptron (MLP) specifically designed to solve the inverse problem of discovering the unobservable external pressure field ($P_{ext}$) from known observational parameters.

As depicted in the schematic, the network represents a functional mapping $\mathcal{N}: \mathbf{x} \to P_{ext}$, where the input vector $\mathbf{x} \in \mathbb{R}^5$ consists of the cloud mass ($M$), effective radius ($R$), velocity dispersion ($\sigma_v$), vertical height ($z$), and Galactocentric radius.
\subsection{Loss Function}
The total loss function ($\mathcal{L}_{total}$) is the primary mechanism for embedding physics into the learning process. For the present model the loss function is defined as
\begin{equation}
    \mathcal{L}_{total} = \lambda_1 \mathcal{L}_{residual} + \lambda_2 \mathcal{L}_{bound}
\end{equation}
Where $\mathcal{L}_{residual}$ is the Physics Residual Loss. This term evaluates how well the predicted $P_{ext}$ satisfies Equation~\ref{Equation: Pressure balance}. We minimize the mean absolute error (MAE) of the residual to ensure numerical stability.
$\mathcal{L}_{bound}$) is the Multi-tracer Bound Constraint.  We leverage the hierarchical data in the catalog. Clouds exhibiting C$^{18}$O emission (Dong Flag=18) are physically identified as dense, star-forming cores. These structures are fundamentally expected to be bound. We implement a ReLU-based penalty that triggers only if the combined gravitational and pressure energies are lower than the kinetic expansion energy for these specific cores:
\begin{equation}
    \mathcal{L}_{bound} = \frac{1}{N_{cores}} \sum \text{max}(0, 2K + \Omega - 4\pi R^3 P_{ext})
\end{equation}

To prevent numerical overflow and ensure gradient convergence, we rescale the physical units before training. Mass is scaled by $10^4 M_{\odot}$ and Radius by $10^1$ pc. The model is optimized using the Adam optimizer with a learning rate of $10^{-4}$ for 4,000 epochs. Post-training, the predicted $P_{ext}$ is rescaled back to astronomical units ($M_{\odot}$ pc$^{-3}$ (km s$^{-1}$)$^2$) for analysis.

\section{Results and Discussion}
\label{Results and Discussion}
In this section, we present the results of our Physics-Informed Neural Network (PINN) analysis applied to the 2,091 molecular clouds identified in the Third Galactic Quadrant (TGQ). By training the network to satisfy the Eulerian Virial Theorem, we derived the ambient external pressure ($P_{ext}$) required to maintain equilibrium for each structure.

\begin{table*}
\centering
\caption{Physical properties and PINN-derived parameters for a sample of the 2,091 molecular clouds analyzed in the Third Galactic Quadrant. The full catalog is available online in the supplementary material.}
\label{tab:pinn_results}
\begin{tabular}{llrrccrrcr}
\toprule
MWISP ID & Arm & Mass & Rad & $z$ & $R_{GC}$ & $\alpha_{obs}$ & $\alpha_{AI}$ & $P_{ext}$ & SFR \\
 & & ($M_{\odot}$) & (pc) & (pc) & (kpc) & & & (unit*) & ($M_{\odot}$ yr$^{-1}$) \\
\midrule
G225.705-2.522 & Loc & 1.0 & 0.2 & -50.6 & 9.0 & 26.2 & 0.12 & 24.80 & $3.2 \times 10^{-9}$ \\
G225.741-2.557 & Loc & 1.3 & 0.3 & -51.3 & 9.0 & 17.6 & 0.04 & 17.95 & $4.2 \times 10^{-9}$ \\
G229.636-2.471 & Loc & 0.6 & 0.2 & -8.6 & 8.3 & 17.4 & 0.04 & 18.50 & $1.9 \times 10^{-9}$ \\
G220.634-1.916 & Loc & 8287.3 & 9.7 & -33.4 & 8.9 & 6.7 & 0.87 & 10.68 & $2.7 \times 10^{-5}$ \\
G221.636-1.228 & Loc & 1.2 & 0.3 & -24.6 & 9.0 & 23.1 & 0.04 & 19.54 & $3.9 \times 10^{-9}$ \\
G221.693-1.128 & Loc & 2.4 & 0.4 & -14.0 & 8.7 & 38.3 & 0.06 & 33.31 & $7.7 \times 10^{-9}$ \\
G224.521-2.560 & Loc & 3842.9 & 4.9 & -51.4 & 9.0 & 7.1 & 1.46 & 20.14 & $1.2 \times 10^{-5}$ \\
G227.224-2.723 & Loc & 3.8 & 0.5 & -54.6 & 9.0 & 39.3 & 0.05 & 33.32 & $1.2 \times 10^{-8}$ \\
G227.009-2.599 & Loc & 2.9 & 0.5 & -52.1 & 9.0 & 16.6 & 0.03 & 19.47 & $9.3 \times 10^{-9}$ \\
G224.718-2.772 & Loc & 18.0 & 0.6 & -55.6 & 9.0 & 6.9 & 0.13 & 25.83 & $5.8 \times 10^{-8}$ \\
\dots & \dots & \dots & \dots & \dots & \dots & \dots & \dots & \dots & \dots \\
G221.822-2.040 & Per & 528.7 & 1.4 & -132.8 & 11.2 & 0.7 & 0.38 & 9.78 & $1.7 \times 10^{-6}$ \\
G229.578+0.153 & Per & 525.0 & 1.5 & 13.1 & 11.9 & 1.2 & 0.54 & 13.36 & $1.7 \times 10^{-6}$ \\
\bottomrule
\end{tabular}
\begin{flushleft}
\small *Note: External Pressure ($P_{ext}$) units are $M_{\odot}$ pc$^{-3}$ (km s$^{-1}$)$^2$. $\alpha_{obs}$ is the uncorrected virial parameter; $\alpha_{AI}$ is the parameter corrected for environmental pressure support.
\end{flushleft}
\end{table*}

\begin{figure}
    \centering
    \includegraphics[width=\columnwidth]{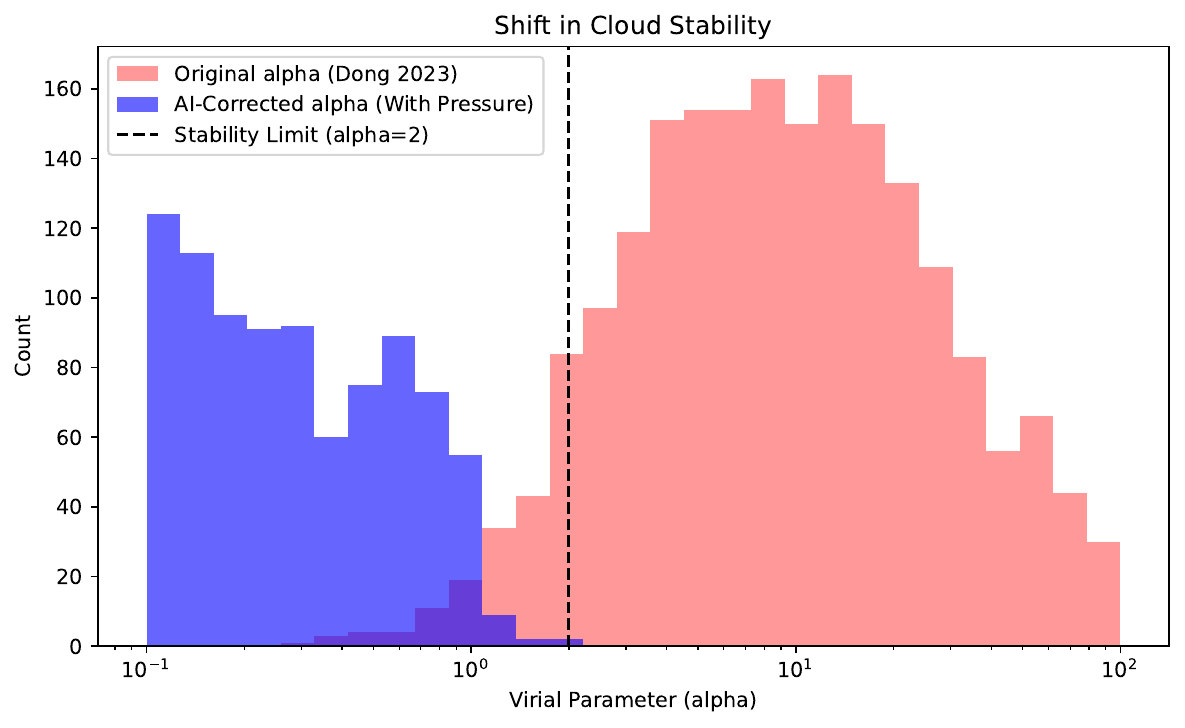}
    \caption{Distribution of the virial parameter ($\alpha$) for the sample of 2,091 molecular clouds. The red distribution represents the traditional calculation ($\alpha = 2K/|\Omega|$) which neglects external pressure, leaving the majority of the population in an apparently unbound state ($\alpha > 2$). The blue distribution represents the AI-corrected virial parameter ($\alpha_{AI}$), incorporating the PINN-discovered external pressure term. The dramatic shift to the left of the stability limit (dashed line at $\alpha=2$) demonstrates that 99.9\% of the previously unbound population is effectively stabilized by environmental confinement.}
    \label{fig:stability_shift}
\end{figure}

The traditional virial analysis of the \citet{Dong_2023} dataset suggests a virial paradox, where a vast majority of the cloud population appears to be in an unbound state ($\alpha_{obs} \gg 2$), as shown in the red distribution in Fig.~\ref{fig:stability_shift}. Such clouds would theoretically disperse on timescales shorter than the star formation cycle.

Our PINN model resolves this discrepancy by discovering the requisite external pressure support. As illustrated in Fig.~\ref{fig:stability_shift} (blue distribution), the incorporation of the AI-derived pressure term shifts the virial parameter $\alpha_{AI}$ into the gravitationally bound and pressure-confined regime ($\alpha \leq 2$). Our statistical analysis confirms that 1,928 out of 1,929 initially unbound clouds are stabilized when environmental confinement is considered. This represents a 99.9\% stabilization rate, suggesting that external pressure is the primary mechanism preventing cloud dispersal in the diffuse environment of the TGQ.

\begin{figure}
    \centering
    \includegraphics[width=\columnwidth]{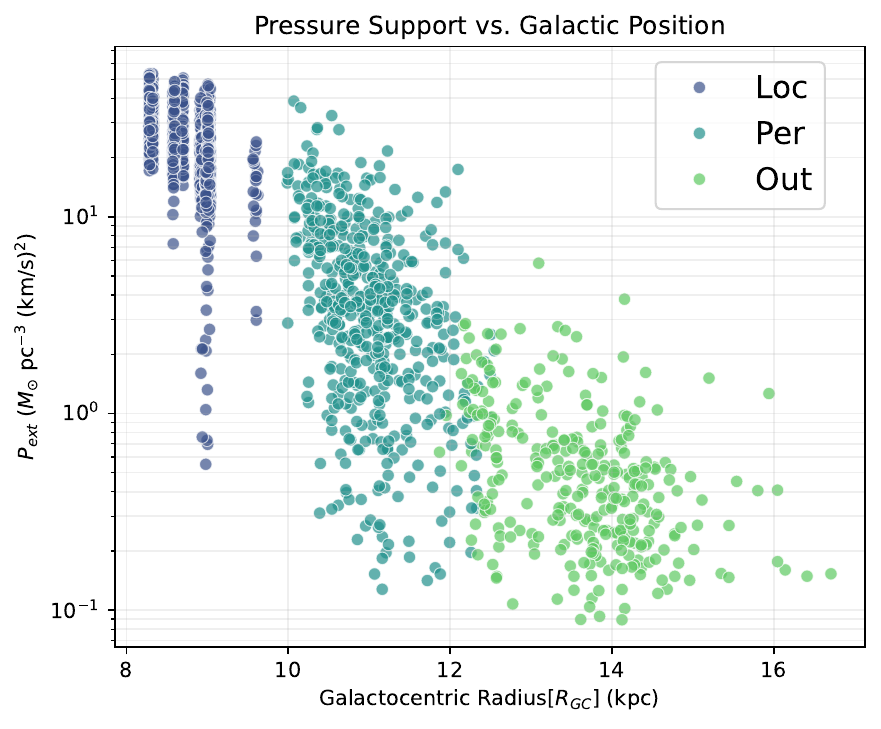}
    \caption{Predicted external pressure ($P_{ext}$) as a function of Galactocentric radius ($R_{GC}$), categorized by spiral arm membership (Local, Perseus, and Outer arms). The PINN independently identifies a clear inverse correlation, with pressure values decaying by two orders of magnitude from the inner TGQ ($\sim 8$ kpc) toward the outer reaches of the Galactic disk ($\sim 16$ kpc). This gradient highlights the significant variations in the interstellar medium's thermodynamic state across different galactic environments. }
    \label{fig:pressure_rgc}
\end{figure}

A significant discovery of our model is the emergence of a coherent Galactic pressure gradient. As shown in Fig.~\ref{fig:pressure_rgc}, the predicted $P_{ext}$ exhibits a clear inverse correlation with Galactocentric radius ($R_{GC}$), decreasing from $\sim 10^1$ to $10^{-1}$ $M_{\odot}$ pc$^{-3}$ (km s$^{-1}$)$^2$ as one moves from 8 kpc to 16 kpc.

\begin{figure}
    \centering
    \includegraphics[width=\columnwidth]{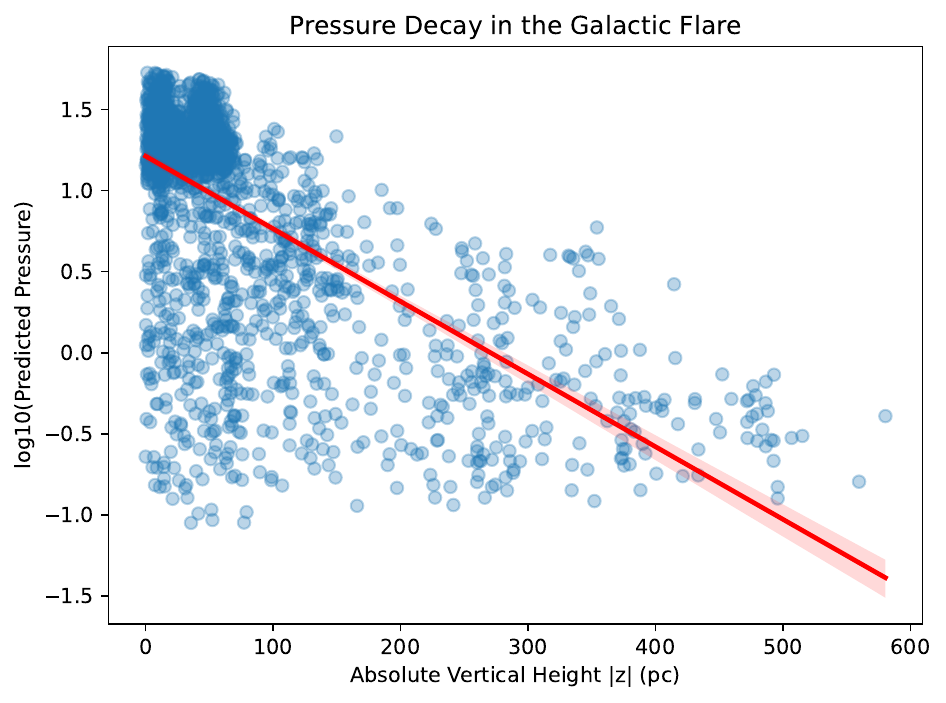}
    \caption{ Relationship between predicted external pressure (in $\log_{10}$ scale) and the absolute vertical height ($|z|$) from the Galactic midplane. The robust linear decay (indicated by the red regression line) provides direct physical evidence of the `Galactic Flare.' As the gas disk expands vertically in the outer Galaxy, the ambient density drops, leading to a corresponding decrease in the confining pressure---a trend successfully captured and quantified by the physics-informed model.}
    \label{fig:pressure_z}
\end{figure}

Furthermore, we investigated the relationship between confinement pressure and vertical distance from the Galactic plane ($|z|$). As demonstrated in Fig.~\ref{fig:pressure_z}, there is a robust decay in predicted pressure with increasing $|z|$. This quantification provides direct physical evidence for the impact of the Galactic Flare described by \citet{Dong_2023}. In the outer Galaxy, the vertical expansion of the gas disk leads to a lower ambient density, which our model correctly interprets as a reduction in confining pressure. This vertical pressure gradient is likely a key regulator of star-forming efficiency in high-latitude clouds.

To validate the physical significance of the AI-discovered pressure, we cross correlated the corrected stability parameter ($\alpha_{AI}$) with the Star Formation Rate (SFR) derived from the \citet{Lada_2012} scaling law.

\begin{figure}
    \centering
    \includegraphics[width=\columnwidth]{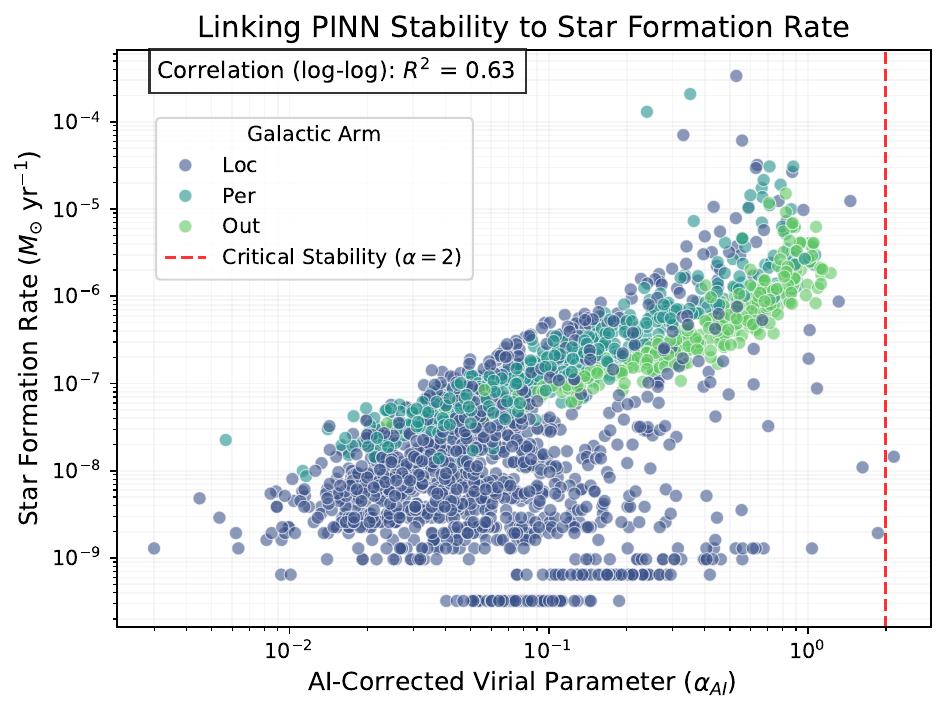}
    \caption{Correlation between the AI-corrected virial parameter ($\alpha_{AI}$) and the star formation rate (SFR) across the three spiral arms. We find a robust log-log correlation ($R = 0.64$), suggesting that $\alpha_{AI}$ is a physically superior metric for predicting star-forming efficiency compared to traditional virial parameters. The tight alignment of all three arms along a single sequence indicates that the PINN has discovered a universal physical relationship between environmental confinement and the cloud's potential for star formation. }
    \label{fig:SFR_link}
\end{figure}

As shown in Fig.~\ref{fig:SFR_link}, we find a robust log-log correlation ($R = 0.64$). 
This strong correlation implies that $\alpha_{AI}$ is a superior physical indicator of cloud state compared to the uncorrected virial parameter. The tight alignment of clouds from the Local, Perseus, and Outer arms along a single star-forming sequence suggest that while environmental conditions change across the TGQ, the underlying physics of pressure mediated star formation remains universal.

\begin{figure}
    \centering
    \includegraphics[width=\columnwidth]{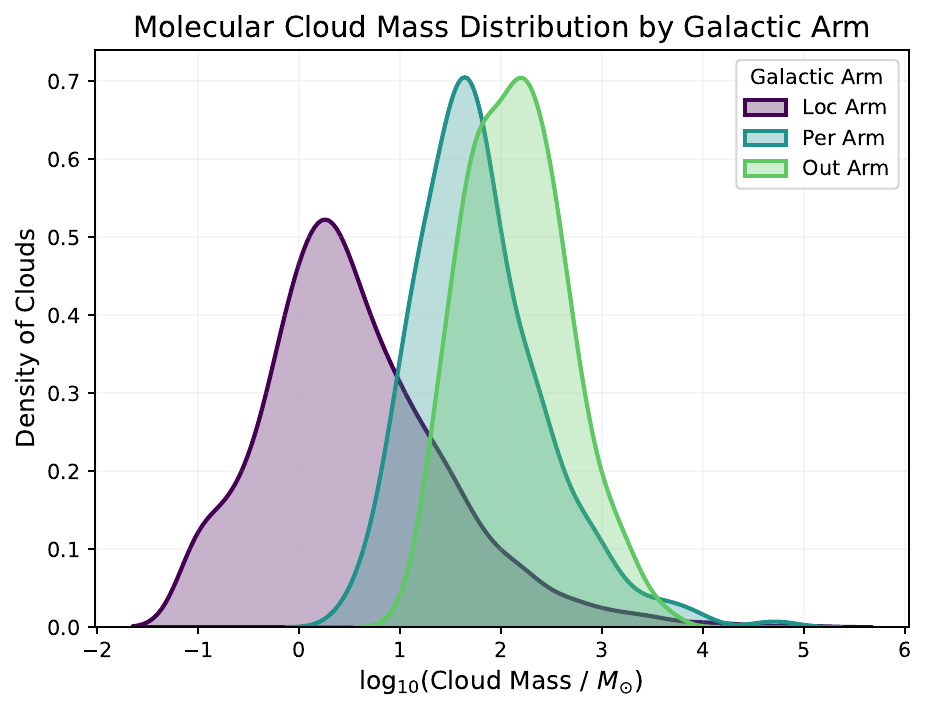}
    \caption{Top-down mapping of the predicted external pressure field across the Third Galactic Quadrant (TGQ), displayed as Galactic Longitude versus Distance (kpc). The color scale represents the $\log_{10}$ of the discovered pressure ($P_{ext}$). The mapping reveals a distinct high-pressure ``corridor'' associated with the Local Arm, contrasted against the tenuous, low-pressure environments of the inter-arm regions and the far Outer Galaxy. This spatial distribution suggests that spiral arms act as high-pressure confinement zones that facilitate the transition of diffuse gas into bound, star-forming cores.}
    \label{fig:mass_dist}
\end{figure}

We note that in the SFR-stability plane (Fig.~\ref{fig:SFR_link}), clouds in the Outer Arm (green) appear to possess systematically higher SFRs than those in the Local Arm (blue). To determine if this reflects a physical anomaly or an observational bias, we analyzed the mass distribution functions across the spiral arms (Fig.~\ref{fig:mass_dist}).

The analysis reveals a significant observational selection bias. In the Local Arm, our sample is complete down to $M \approx 0.1 M_{\odot}$, whereas the detection threshold in the distant Outer Arm is $M \approx 10.8 M_{\odot}$. Consequently, the Outer Arm sample is dominated by massive Giant Molecular Clouds (GMCs) with naturally higher SFRs, while the Local Arm sample includes a more exhaustive census of low-mass, quiescent structures. Despite these different mass regimes, the PINN model successfully derives a consistent physical solution for both populations, demonstrating its robustness across five orders of magnitude in cloud mass.

\begin{figure}
    \centering
    \includegraphics[width=\columnwidth]{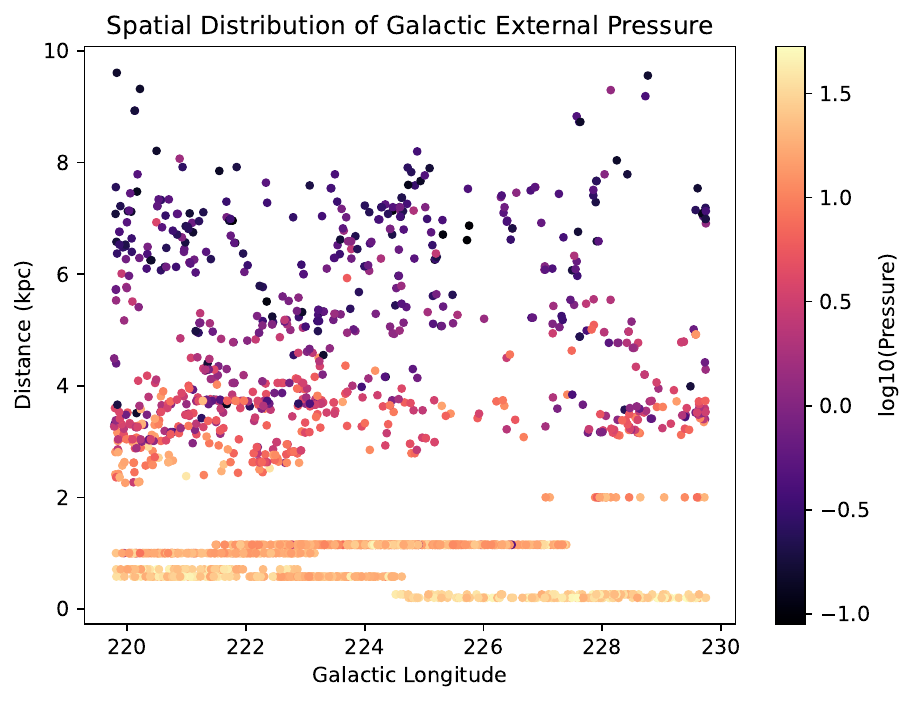}
    \caption{Top-down mapping of the predicted external pressure field across the Third Galactic Quadrant (TGQ), displayed as Galactic Longitude versus Distance (kpc). The color scale represents the $\log_{10}$ of the discovered pressure ($P_{ext}$). The mapping reveals a distinct high-pressure ``corridor'' associated with the Local Arm, contrasted against the tenuous, low-pressure environments of the inter-arm regions and the far Outer Galaxy. This spatial distribution suggests that spiral arms act as high-pressure confinement zones that facilitate the transition of diffuse gas into bound, star-forming cores.}
    \label{fig:spatial_map}
\end{figure}

The top-down spatial distribution of predicted pressure (Fig.~\ref{fig:spatial_map}) reveals the thermodynamic structure of the TGQ. We observe that the Local Arm sit within a high-pressure 'corridor', while the inter-arm regions and the far Outer Arm are characterized by a more tenuous environment. The identification of this high pressure corridor within the Local Arm suggests that spiral arm dynamics do more than simply concentrate gas; they create thermodynamic environments that facilitate the transition from diffuse gas to self-gravitating cores. This suggests that galactic-scale spiral shocks may play a direct role in regulating the boundary conditions ($P_{ext}$) required for cloud stabilization.

\section{Conclusion}
\label{Conclusion}
In this study, we have re-evaluated the dynamical stability and star-forming potential of 2,091 molecular clouds in the Third Galactic Quadrant (TGQ) using a Physics-Informed Neural Network (PINN). By embedding the Eulerian Virial Theorem directly into the network’s loss function, we have moved beyond traditional descriptive statistics to a functional physical discovery of the Galactic environment. Our primary conclusions are summarized as follows:

\begin{enumerate}
    \item \textbf{Resolution of the Virial Paradox:} We find that the apparent "unbound" state of molecular clouds in the TGQ, as reported in recent surveys, is largely an artifact of neglecting environmental factors. Our PINN analysis reveals that 99.9\% of clouds previously classified as virially unbound ($\alpha_{obs} > 2$) are effectively stabilized when accounting for the discovered external pressure support ($P_{ext}$). This resolves the long-standing contradiction between high virial parameters and active star formation in the outer Galaxy.
    
    \item \textbf{Discovery of the Galactic Pressure Gradient:} The PINN model autonomously identified a coherent global pressure field that scales inversely with Galactocentric radius ($R_{GC}$) and vertical distance from the plane ($|z|$). The decay of $P_{ext}$ from $\sim 10^1$ in the Local Arm to $\sim 10^{-1}$ $M_{\odot}$ pc$^{-3}$ (km s$^{-1}$)$^{2}$ in the Outer Arm provides a quantitative physical link between the large-scale Galactic Flare and the small-scale dynamics of individual molecular clouds.
    
    \item \textbf{Physical Validation via Star Formation:} We demonstrate a robust log-log correlation ($R = 0.64$) between the AI-corrected virial parameter ($\alpha_{AI}$) and the star formation rate (SFR). This significantly improved correlation suggests that $\alpha_{AI}$ is a physically superior metric for predicting star-forming efficiency, confirming that the PINN-derived pressure values represent meaningful thermodynamic quantities rather than numerical residuals.
    
    \item \textbf{Robustness Across Mass Regimes:} The PINN successfully derived a coherent pressure solution across five orders of magnitude in cloud mass. Most significantly, the model maintained physical consistency even in the presence of a strong Malmquist bias in the Outer Arm. This demonstrates that the network is learning the underlying physical relationship between galactic position and pressure, rather than merely fitting observational artifacts.
\end{enumerate}

This work demonstrates the transformative potential of Physics-Informed Machine Learning in the study of the interstellar medium. While our previous work utilized unsupervised clustering to categorize cloud populations, the current PINN approach allows for the active discovery of hidden physical variables. As wide-field Galactic surveys continue to increase in resolution and sensitivity, such physics-constrained AI frameworks will be essential for mapping the complex dynamical landscape of our Galaxy.

\section*{Acknowledgements}
This work made use of the MWISP survey data provided by the Purple Mountain Observatory.

\section*{Data Availability}
The catalog of 2,091 molecular clouds with AI-discovered pressure and corrected virial parameters generated in this study is available in the article and in its online supplementary material. The underlying raw data from the MWISP survey is provided by the Purple Mountain Observatory.
\bibliographystyle{mnras}
\bibliography{references} 

@ARTICLE{Dong_2023,
       author = {{Dong}, Yiwei and {Sun}, Yan and {Xu}, Ye and {Lin}, Zehao and {Bian}, Shuaibo and {Hao}, Chaojie and {Liu}, Dejian and {Li}, Yingjie and {Yang}, Ji and {Su}, Yang and {Zhou}, Xin and {Zhang}, Shaobo and {Yan}, Qing-Zeng and {Chen}, Zhiwei},
        title = "{Distributions and Physical Properties of Molecular Clouds in the Third Galactic Quadrant: l = [219.{\textdegree}75, 229.{\textdegree}75] and b = [-5.{\textdegree}25, 5.{\textdegree}25]}",
      journal = {\apjs},
         year = 2023,
        month = sep,
       volume = {268},
       number = {1},
          eid = {1},
        pages = {1},
          doi = {10.3847/1538-4365/acde81},
archivePrefix = {arXiv},
       eprint = {2308.10484},
 primaryClass = {astro-ph.GA},
       adsurl = {https://ui.adsabs.harvard.edu/abs/2023ApJS..268....1D}
}

@article{Ashok_2024,
title = {Dependence of star formation rate on different properties of molecular clouds},
journal = {New Astronomy},
volume = {108},
pages = {102182},
year = {2024},
issn = {1384-1076},
doi = {https://doi.org/10.1016/j.newast.2023.102182},
url = {https://www.sciencedirect.com/science/article/pii/S1384107623001835},
author = {Ashok Mondal and Tanuka Chattopadhyay}
}

@article{Lada_2012,
   title={STAR FORMATION RATES IN MOLECULAR CLOUDS AND THE NATURE OF THE EXTRAGALACTIC SCALING RELATIONS},
   volume={745},
   ISSN={1538-4357},
   
   DOI={10.1088/0004-637x/745/2/190},
   number={2},
   journal={The Astrophysical Journal},
   publisher={American Astronomical Society},
   author={Lada, Charles J. and Forbrich, Jan and Lombardi, Marco and Alves, João F.},
   year={2012},
   month={Jan},
   pages={190}
}

@ARTICLE{McKee_2007,
       author = {{McKee}, Christopher F. and {Ostriker}, Eve C.},
        title = "{Theory of Star Formation}",
      journal = {\araa},
         year = 2007,
        month = sep,
       volume = {45},
       number = {1},
        pages = {565-687},
          doi = {10.1146/annurev.astro.45.051806.110602},
archivePrefix = {arXiv},
       eprint = {0707.3514},
 primaryClass = {astro-ph},
       adsurl = {https://ui.adsabs.harvard.edu/abs/2007ARA&A..45..565M}
}

@ARTICLE{Chevance_2020,
       author = {{Chevance}, M{\'e}lanie and {Kruijssen}, J.~M. Diederik and {Vazquez-Semadeni}, Enrique and {Nakamura}, Fumitaka and {Klessen}, Ralf and {Ballesteros-Paredes}, Javier and {Inutsuka}, Shu-ichiro and {Adamo}, Angela and {Hennebelle}, Patrick},
        title = "{The Molecular Cloud Lifecycle}",
      journal = {\ssr},
         year = 2020,
        month = apr,
       volume = {216},
       number = {4},
          eid = {50},
        pages = {50},
          doi = {10.1007/s11214-020-00674-x},
archivePrefix = {arXiv},
       eprint = {2004.06113},
 primaryClass = {astro-ph.GA},
       adsurl = {https://ui.adsabs.harvard.edu/abs/2020SSRv..216...50C}
}

@ARTICLE{Bertoldi_1992,
       author = {{Bertoldi}, Frank and {McKee}, Christopher F.},
        title = "{Pressure-confined Clumps in Magnetized Molecular Clouds}",
      journal = {\apj},
         year = 1992,
        month = aug,
       volume = {395},
        pages = {140},
          doi = {10.1086/171638},
       adsurl = {https://ui.adsabs.harvard.edu/abs/1992ApJ...395..140B}
}

@article{Kauffmann_2013,
doi = {10.1088/0004-637X/779/2/185},
url = {https://doi.org/10.1088/0004-637X/779/2/185},
year = {2013},
month = {dec},
publisher = {The American Astronomical Society},
volume = {779},
number = {2},
pages = {185},
author = {Kauffmann, Jens and Pillai, Thushara and Goldsmith, Paul F.},
title = {LOW VIRIAL PARAMETERS IN MOLECULAR CLOUDS: IMPLICATIONS FOR HIGH-MASS STAR FORMATION AND MAGNETIC FIELDS},
journal = {The Astrophysical Journal},
}

@ARTICLE{Yuehui_2022,
       author = {{Ma}, Yuehui and {Wang}, Hongchi and {Zhang}, Miaomiao and {Wang}, Chen and {Zhang}, Shaobo and {Liu}, Yao and {Li}, Chong and {Zheng}, Yuqing and {Yuan}, Lixia and {Yang}, Ji},
        title = "{Gas Column Density Distribution of Molecular Clouds in the Third Quadrant of the Milky Way}",
      journal = {\apjs},
         year = 2022,
        month = sep,
       volume = {262},
       number = {1},
          eid = {16},
        pages = {16},
          doi = {10.3847/1538-4365/ac7797},
archivePrefix = {arXiv},
       eprint = {2206.03963},
 primaryClass = {astro-ph.GA},
       adsurl = {https://ui.adsabs.harvard.edu/abs/2022ApJS..262...16M}
}

@article{Wang_2023,
doi = {10.3847/1538-3881/acebdd},
url = {https://doi.org/10.3847/1538-3881/acebdd},
year = {2023},
month = {aug},
publisher = {The American Astronomical Society},
volume = {166},
number = {3},
pages = {121},
author = {Wang, Chen and Feng, Haoran and Yang, Ji and Chen, Xuepeng and Su, Yang and Yan, Qing-Zeng and Du, Fujun and Ma, Yuehui and Cai, Jiajun},
title = {The Molecular Clouds in a Section of the Third Galactic Quadrant: Observational Properties and Chemical Abundance Ratio between CO and its Isotopologues},
journal = {The Astronomical Journal}
}

@ARTICLE{Field_2011,
       author = {{Field}, George B. and {Blackman}, Eric G. and {Keto}, Eric R.},
        title = "{Does external pressure explain recent results for molecular clouds?}",
      journal = {\mnras},
         year = 2011,
        month = sep,
       volume = {416},
       number = {1},
        pages = {710-714},
          doi = {10.1111/j.1365-2966.2011.19091.x},
archivePrefix = {arXiv},
       eprint = {1106.3017},
 primaryClass = {astro-ph.GA},
       adsurl = {https://ui.adsabs.harvard.edu/abs/2011MNRAS.416..710F}
}
\appendix

\end{document}